# Substance graphs are optimal simple-graph representations of metabolism

Petter Holme[1,2] and Mikael Huss[3]

[1]*Computational Biology, Royal Institute of Technology, 100 44 Stockholm, Sweden*
[2]*Department of Physics, Umeå University, 901 87 Umeå, Sweden*
[3]*Systems Biology, Genome Institute of Singapore, 138 672 Singapore, Singapore*

One approach to studying the system-wide organization of biochemistry is to use statistical graph theory. Even in such a heavily simplified method, which disregards most of the dynamic aspects of biochemistry, one is faced with fundamental questions, such as how the chemical reaction systems should be reduced to a graph retaining as much functional information as possible from the original reaction system. In such graph representations, should the edges go between substrates and products, or substrates and substrates, or both? Should vertices represent substances or reactions? Different definitions encode different information about the reaction system. In this paper we evaluate four different graph representations of metabolism, applied to data from different organisms and databases. The graph representations are evaluated by comparing the overlap between clusters (network modules) and annotated functions, and also by comparing the set of identified currency metabolites with those that other authors have identified using qualitative biological arguments. We find that a "substance network," where all metabolites participating in a reaction are connected, is relatively better than others, evaluated both with respect to the functional overlap between modules and functions and to the number and identity of identified currency metabolites.

## I. INTRODUCTION

Metabolism, the set of all chemical processes in an organism that are necessary for the maintenance of life, can be studied at different levels — from the quantum chemistry of reactions via small reaction pathways and larger feedback loops to system-wide organization. For the smaller-scale problems, sets of reactions are often modeled by systems of ordinary differential equations (ODEs), either in order to perform explicit simulations or to make use of steady-state properties of the systems such as in "flux mode analysis" (17). However, to investigate the system-wide, large-scale organization for whole organisms, such modeling frameworks become obstructively complex. An alternative paradigm is graph methods, where the chemical reaction system is reduced to a graph of nodes (or *vertices*) pairwise connected by links (*edges*). In this paper, we define a *metabolic network* as any *simple graph* (unweighted, undirected graph without multiple edges or self-edges) derived from a metabolic reaction system. In the process of constructing a metabolic network, one has to discard much of the existing information about reaction systems (information about reaction coefficients, localization etc.). In return, one obtains an object of study — a graph — for which numerous methods exists to characterize its large-scale organization.

As illustrated in Fig. 1, there are several ways to represent chemical reaction systems, such as an organism's set of metabolic reactions, as simple graphs. The most common way, Fig. 1b, is to link substrates to products in a *substrate–product network* (sometimes called just "substrate network" (22)). Furthermore, one can consider linking products to products and substrates to substrates into what we call a *substrate–substrate network* (Fig. 1c), or linking all metabolites participating in the same reaction with each other into a *substance network* (Fig. 1d). As a final representation, instead of linking all substrates participating in the same reaction, one can make a *reaction network* (Fig. 1e) of reactions connected if they have a substance in common. These different representations can potentially capture different aspects of the reaction

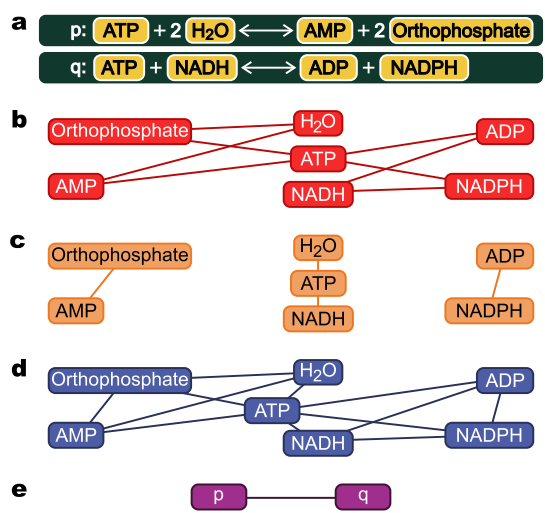

FIG. 1 Panel **a** shows two reactions that together form a minimal reaction system. **b** displays the substrate–product network derived from the reaction system in **a**. Here, the vertices are chemical substances that are connected if they are substrate and product of a reaction. **c** shows the substrate–substrate network where substances are connected if they react with each other. **d** illustrates the substance network where substances are connected if the occur in the same reaction. **e** shows the reaction network derived from **a**, where two reactions are connected if they share a substance.

system — the graph distances in a substrate–product network reflect the number of reactions atoms traverse on the way between two substances, and so on. For specific questions about the system, one of these representations might be more useful than the others. Most questions appropriate for a graph-based approach, however, hinge on the assumption that the wiring mirrors the functional organization of metabolism. To evaluate the different representations, we use three empirical criteria for investigating how well the networks capture biological functionality. First, we investigate the assumption (1) that a tightly knit group of metabolites are relatively likely to be involved in the same biological processes, and investigate how well network clusters match annotated functions. Second, we investigate the quality of *currency metabolites* (ubiquitous substances, like water and carbon dioxide, participating in a large number of reactions; a more precise definition follows below) derived from the network. Finally, making the assumptions that metabolic networks show modular structure, we require that the network representation should yield a clear modularity after currency metabolites have been removed.

## II. RESULTS

In this paper we use the reaction systems of ten datasets (eight different organisms from the KEGG and BiGG databases, accessed April 2008). Sizes and basic statistics of the derived metabolic networks are given in Appendix 1. We use a previously described algorithm (6) to decompose the networks into modules. This algorithm essentially maximizes network modularity by successively removing high-degree metabolites. Its output is twofold: on one hand, it returns a set of clusters (or modules), and on the other hand, it identifies a set of currency metabolites. The resulting modules should be enriched in molecules having similar annotation, and the resulting currency metabolites should correspond to substances that have been described as currency metabolites by previous authors. These will be our main criteria for evaluating the different graph representations. Also, Refs. (4; 5; 7; 11; 15; 21; 23) argue that network modularity is a principle of metabolic organization. Thus, we also expect that a good graph representation that the modular structure should be clear after the deletion of the currency metabolites.

### A. Functional matching

What is the nature of the dense clusters of non-currency metabolites detected by our decomposition algorithm? The edges in all four representations represent some kind of functional relationship between the connected vertices. Although these functional relationships are slightly different, one can expect the vertices of a cluster to have a stronger functional coupling to each other than to vertices of other clusters. In other words, a cluster should be functionally relatively independent, which is one aspect of the biological notion of a module. To investigate the relationship between clusters and modules, we measure a matching score (defined in the Methods section), displayed in Fig. 2a. A positive value of the matching score represents an overlap greater than expected from a situation where the annotated functions were randomly spread out over the vertices (i.e., relative to a randomized set of the functions). Positive values of the functional overlap cannot be larger than unity. Examining the scores for different organisms in Fig. 2a, we notice that all scores are significantly postive, meaning that there is a tendency for functionality to be concentrated to the network clusters. Comparing the different representations, we see that the reaction networks seem to be particularly poor in capturing the functional aspects of the reaction system. The other three representations appear to have similar performance, with the substance network having especially good performance in the human BiGG and the *M. genitalium* data sets.

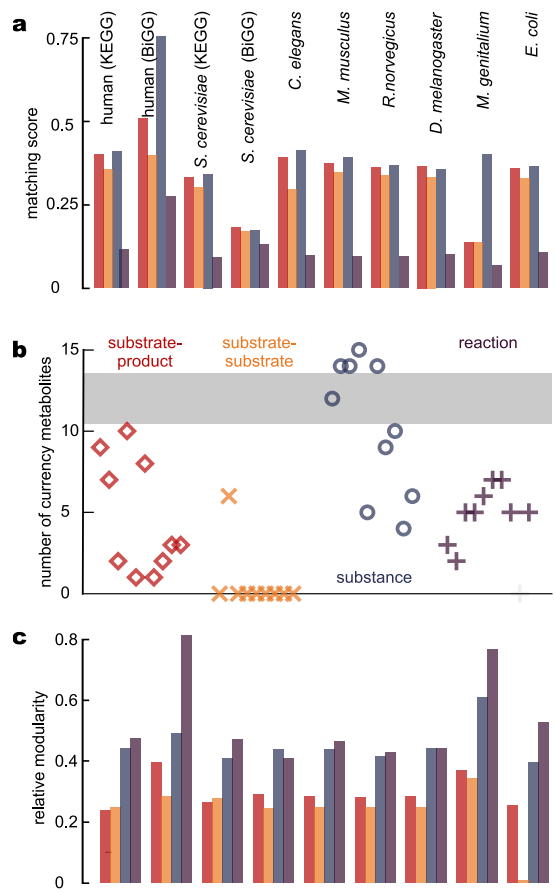

FIG. 2 Panel **a** shows the functional matching score for the different data sets and graph representations. **b** displays the number of currency metabolites. The shaded area represent the number of currency metabolites suggested by other authors. Panel **c** shows the relative modularity of the networks, the order of the data corresponding to panel **a**. The color coding in panels **a** and **c** is the same as in panel **b**.

### B. Currency metabolites

Hubs in biomolecular networks are usually abundant molecular species (6) which do not put much constraint on the metabolic flows. Accordingly, they tend not to have much regulatory (or other higher-order) functionality. The highest-degree vertices typically connect different network clusters, thus blurring the modular structure of the network. This property (illustrated by the human substance network from the KEGG database in Fig. 3) and the high degree are the two components in our operational definition of currency metabolites (5): *If vertices are deleted from the network in order of highest degree, then the set of currency metabolites is the set of vertices that, if deleted, gives the highest relative modular-*

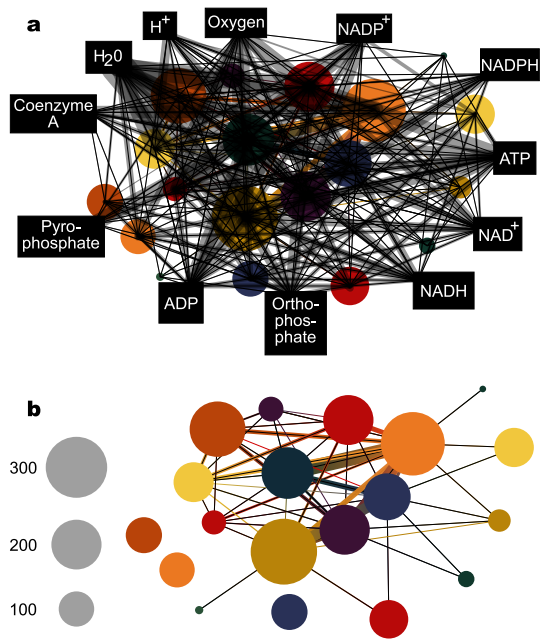

FIG. 3 Illustration of the network structural role of currency metabolites. The figure is based on the human substance network from the KEGG database. The modules are indicated in different colors (areas of the circles are proportional to the size of the clusters). The width of the lines are proportional to the number of edges leading between two different clusters, or from a currency metabolite to a cluster. In panel **b**, the currency metabolites are removed from the picture and the modular structure is more visible.

TABLE I A comparison between currency metabolites detected by our algorithm (all detected in at least one of eight organisms using substance networks) and currency metabolites defined by Wagner and Fell (20), Schuster *et al.* (18), and Ma and Zeng (10). Only metabolites which were considered currency metabolites in at least two of the four studies are shown. In Ref. (20), the analysis was done in two versions, one excluding and one including the six first metabolites (ATP to NADH). In Refs. (10) and (18), the sets of currency metabolites were considered context-dependent and thus not completely fixed, hence these lists are approximations. A list of currency metacolites for all studied organisms and graph types in this work can be found in Appendix 2.

| this work | Wagner & Fell | Schuster *et al.* | Ma & Zeng |
|---|---|---|---|
| ATP | ATP | ATP | ATP |
| ADP | ADP | ADP | ADP |
| NADPH | NADPH | NADPH | NADPH |
| NADP$^+$ | NADP$^+$ | NADP$^+$ | NADP$^+$ |
| NAD$^+$ | NAD$^+$ | NAD$^+$ | NAD$^+$ |
| NADH | NADH | NADH | NADH |
| P$_i$ | P$_i$ | P$_i$ | P$_i$ |
| PP$_i$ | PP$_i$ | PP$_i$ | PP$_i$ |
| CO$_2$ | CO$_2$ | CO$_2$ | CO$_2$ |
| H$_2$O | | H$_2$O | H$_2$O |
| | NH$_3$ | NH$_3$ | NH$_3$ |
| | SO$_4$ | SO$_4$ | SO$_4$ |
| H$^+$ | | H$^+$ | |
| O$_2$ | | | O$_2$ |

*ity.* Note that this definition does not specify the graph representation. Early papers (10; 18; 20) have defined currency metabolites based on biological reasoning. Although these sets of currency metabolites are somewhat disparate (Table I), we recognize the relevance of these biological arguments. A quality criterion of the graph representations would then be that the currency metabolites defined by these representations should have considerable overlap with these other studies. A first check is that the number of currency metabolites is of the same order, between 11 and 14, as proposed in the other studies (10; 18; 20). In Fig. 2b we plot the number of currency metabolites for ten organisms of various complexity, from two different databases. None of the representation produce a very consistent set of currency metabolites. This is not expected either — consider, for example, oxygen, which appears as a currency metabolite in the human metabolic network according to our algorithm; in the metabolism of anaerobic bacteria, it will most likely not be a currency metabolite (cf. Ref. (16). The representation whose number of currency metabolites agrees best with the range 11–14 in humans is the substance network.

## C. Modular structure

In Fig. 2c we plot the relative modularity $\Delta$ in a fashion corresponding to Fig. 2a. In general, the substance and reaction networks show the highest $\Delta$-value. This suggests that these representations may be superior in capturing modular structure of metabolic networks. $\Delta$ (defined in Methods) is a modification of Newman and Girvan's $Q$-modularity (14) for the algorithm to detect currency metabolites. $Q$ is known to be biased by size and degree sequence (3; 8). While $\Delta$ removes enough of these effects to work in the currency metabolite algorithm (where we only compare the marginal effect on $\Delta$ on the removal of vertices), the size differences between the different representations might affect the conclusions. If a cautious conclusion is to be drawn, the large difference between $\Delta$ of the substance network compared with the substrate–product and substrate–substrate representations suggests that the substrate network representation is likely to be better at capturing the modular structure of the network, although the reaction network (which performs poorly for the other two criteria) is even better than the substance network. This high-modularity criterion is the weakest one, as it does not relate to any external information. Nevertheless, the outcome of the test is at least not an argument against the substance network representation.

## D. Functional organization — A continuum between a core and a modular periphery

From a biological viewpoint, maybe the most important conclusion from Fig. 2a is that the functions are concentrated to the network clusters — $\mu$ is significantly larger than zero for all graph representations and organism. But how is the functionality distributed among the clusters? One description of metabolism is that currency metabolites form the core of metabolism and the higher order functionality located to the clusters. There are, however, studies indicating that this picture is not completely dichotomous (4; 23), as some of the clusters bridge the currency metabolites and modules of higher functionality. Another type of modular organization with central and peripheral modularity layers, separated by a zone of depressed modularity, has recently been described in protein domain networks (19) but has not yet been investigated in metabolic networks. Here we will argue for an organization characterized by a core — the currency metabolites — and clusters of increasing specificity. To investigate this picture (for the substance network), in Fig. 4 we plot the contribution from the cluster to $\mu$ as a function of the fraction of all edges from the cluster to currency metabolites. Our picture corresponds to a negative relationship between these two quantities. Indeed, there is an overall decreasing trend, but not a very strong correlation. One reason, we believe, is that biases are introduced by incompleteness of the functional annotation. For example, the average number of functions per metabolite in the respective clusters (indicated by color) seems correlated to high functional overlap. This may be an artifact of more well-characterised compounds having more known functions and interactions. We believe that for future datasets with more precise functional assignment, the clusters in the lower left corner (usually small clusters with little association to either functional overlap or currency metabolites) will move upwards. This picture holds for other organisms, see the Appendix 3. Indeed, the human network of Fig. 4 is not illustrating our picture most clearly.

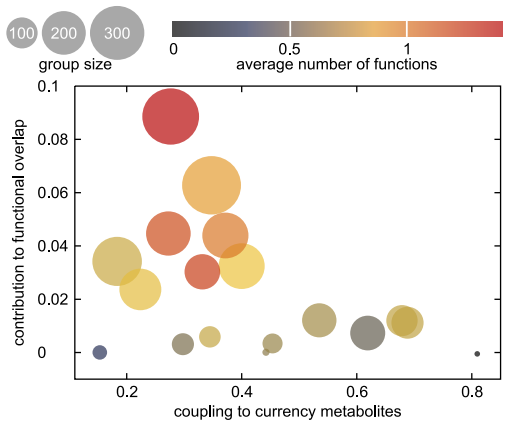

FIG. 4  A figure illustrating the relation between functionality and network-cluster structure of the human substance network from the KEGG database. For all detected network clusters, we plot their contribution to the functional matching score as a function to the fraction of edges leading to currency metabolites. The sizes of the circles are proportional to the sizes of the clusters. Color indicates the average number of annotated functions per metabolite.

### E.  Discussion

In this paper we have discussed four different types of graph representations of metabolism — substrate–product networks, substrate–substrate networks, substance networks and reaction networks. The first three representations are more or less the only ways, using simple rules and available database information, to reduce a reaction system to a graph of interlinked substances. For reaction graphs one can imagine other varieties (e.g., connecting reactions sharing substrates), but in this work we consider reaction graphs mostly to illustrate the possibility.

We use quantitative arguments for assessing which one of these representations that is the most informative about the functional organization of metabolic reaction networks. The criteria are that the network modules should have a high overlap with annotated functions, identified currency metabolites should be similar to those of other studies, and the modular structure should be clear. With these criteria, the substance graph representation emerges as a relatively clear winner. The other representations may, however, be preferable for more specific questions. For instance, a substrate–product graph stresses the interconversion between substances, making it possible to trace paths between different substances. Substrate–substrate graphs highlight reactivity between metabolites. A substance graph embodies information about spatial and temporal co-localization of substances, while the reaction networks show how the reactions are connected. An investigator may be able to select a suitable representation depending on the intended usage of the network, be it bottleneck identification, lethality prediction, or one of any number of possible applications.

Analyzing the substrate networks, we can describe the large-scale functional organization of metabolism as centered around a core of currency metabolites with other clusters forming a periphery of gradually increasing functional specificity, decreasing multi-functionality and decreasing coupling to currency metabolites. A full verification of this picture has to wait for future functional annotations of higher precision.

In future work, it would be interesting to compare the different graph representations in a dynamic setting with some process acting over the vertices, or in a task-oriented setting where the networks are used to address a specific problem. However, we believe that our work presented here, based on simple, static graphs (for example, relating metabolites cellular and pathogenic functions (9)) is a necessary first step to understand the basic differences between the graph types.

## III.  METHODS

### A.  Network modularity and currency metabolites

Real-world complex networks, including metabolic networks, can be described as having both randomness and some regularities, or *network structure*. The network structure contains information both about how dynamic systems associated with the network behave, and about the evolutionary history of the network itself. A key to accessing this information is to design quantities to measure network structure. One such type of network-structural quantities are measures of network modularity — how well a graph can be decomposed into sub-networks that are densely connected within and sparsely connected between each other. In Ref. (14) a measure of modularity of a partition of a graph into subgraphs was proposed:

$$Q = \sum_i \left[ e_{ii} - \left( \sum_j e_{ij} \right)^2 \right],$$ (1)

where the sum is over subgraphs and $e_{ij}$ is the fraction of edges that leads between vertices of cluster $i$ and $j$. The term $\left( \sum_j e_{ij} \right)^2$ is the expected value of $e_{ii}$ in a random multigraph (without any correlations). One way of detecting (or defining) clusters in networks is to maximize $Q$ over all partitions, of all sizes. This is a computationally hard optimization problem (2) that a large number of papers has proposed heuristics for solving (see Ref. (13) and references therein). We use a recent and competitive algorithm proposed in Ref. (13).

The maximal value $\hat{Q}$ of $Q$ for all partitions is a prototype measure of the modularity of a network. However, as hinted above, network structure needs to be measured relative to a null model. The most common null-model for metabolic networks is the ensemble $\mathcal{G}(G)$ of random, simple graphs with the (only) condition that they should have the same degree sequences (set of degrees, i.e. number of neighbors) as the original network $G$. A measure of effective, relative modularity is thus:

$$\Delta(G) = \hat{Q}(G) - \langle \hat{Q}(G') \rangle_{G' \in \mathcal{G}(G)},$$ (2)

where the angular brackets denote average over $\mathcal{G}(G)$. The common way of sampling $\mathcal{G}(G)$ (and the method we employ) is a resampling technique, randomly rewiring the edges of the original network (12). In this work we use averages over 100 independent samples of this ensemble.

Like the degree of metabolic networks (Appendix 4), the abundance of metabolites has a very broad distribution (6). Currency metabolites are typically abundant molecules that react with a variety of other metabolites having many different functions and (by the above-mentioned assumption that metabolism has functional modules corresponding to network clusters) belonging to many different functional modules. This means that, in addition to having a large degree, currency metabolites will also effectively lower the network modularity; see the Appendix 2. These two properties combined are the motivation for above stated definition of currency metabolites (if vertices are deleted from the network in order of highest degree, then the set of currency metabolites is the set of vertices that, if deleted, gives the highest relative modularity). This is, to our knowledge, the only definition of currency metabolites from measurable properties of the metabolites. In this paper we add the criterion that if $\Delta$ decreases below its original value, we break the iterations. This is done to speed up the calculation, and does not affect the output for the (human KEGG and BiGG) networks we test this property for. When we apply this algorithm to the reaction network (where the vertices represent reactions), we still delete substances (in order of the number of reactions they participate in), but measure the modularity change of the reaction network. A C implementation of the currency-metabolite detection algorithm can be found at `http://www.csc.kth.se/~pholme/curr/`.

## B. Functional matching

Metabolites in KEGG are annotated with one or several pathways which we equate with functions. In BiGG, similar functional (pathway) annotations are instead assigned to reactions, but one can indirectly assign functions to a metabolite by simply selecting all functions listed for reactions where the metabolite participates. Every metabolite $i \in V$ is thus associated with a set of functions $f_i \in F$ ($F$ is the set of all functions). We can use this to evaluate the assumption that network modules correspond to biological functions. Let $\phi_{CF}(c, f)$ denote the fraction of vertices with function $f$ belonging to cluster $c \in C$ ($C(G)$ is the set of detected clusters of $G$); let $\phi_F(f)$ denote the fraction of vertices with function $f$; and let $\phi_C(c)$ denote the fraction of vertices in cluster $c$. Then, if the functions were randomly distributed, in an infinite system, the expectation value of $\phi_{CF}$ would be $\phi_C(c)\phi_F(f)$. From this we derive a prototypical score function for the match between network modules and metabolite function:

$$\nu = \sum_{c \in C} \sum_{f \in F} |\phi_{CF}(c, f) - \phi_C(c)\phi_F(f)|,$$ (3)

where $|\cdots|$ denotes absolute values of numbers and cardinality (number of elements) of sets. High values of $\nu$ mean that functions are concentrated to network clusters. If the number of vertices and the sizes of clusters go to infinity, $\nu = 0$ signals neutrality. Note that this definition does not require the vertex set to be partitioned into a set of distinct clusters (similarly it allows several functions to be assigned to one vertex). When we apply the currency metabolite definition algorithm to the product network, we still delete vertices (and all the reactions they participate in), but define the currency metabolites with respect to the modularity of the reaction network. This means that the network clusters will be sets of reactions. When calculating $\nu$ for the reaction network, we let a metabolite belong to the network clusters of all the reactions in which it participates. However, $\nu$ does not capture all the aspects we desire — since fluctuations give a positive contribution to $\nu$ (due to the absolute values), the finite sizes of the reaction systems will give a positive bias. To make zero represent neutrality, we rather measure

$$\mu = \frac{\nu - \langle \nu' \rangle}{\nu^* - \langle \nu' \rangle},$$ (4)

where $\nu'$ is a random configuration of functions with the properties that $F$ and $\phi_F(f)$ (for all $f \in F$) is the same as in the original system, that no vertex is assigned the same function more than once, and $\langle \cdots \rangle$ denotes the average over 100 samples of this ensemble, and $\nu^*$ represents the ensemble's maximal value.

## Acknowledgments

PH acknowledges economic support from the Swedish Foundation for Strategic Research. The authors thank Lei-Han Tang and Jing Zhao for insightful comments.

Appendix 1:
**Network sizes**

**Substance graphs are optimal simple-graph representations of metabolism**

Petter Holme and Mikael Huss

---

In this file we present the sizes of the networks and give a brief explanation to why the reaction networks are denser than other networks.

## Some fundamental quantities

Table of some fundamental quantities — the number of vertices $N$, edges $M$, currency metabolites $c$ and relative modularity $\Delta$ — for the various types of networks, for different organisms and different databases (KEGG, unless otherwise stated). $\Delta$ and $c$ are compared visually in Fig. 3 of the paper.

| organism | $N$ | substr.–prod. | | | substr.–substr. | | | substance nwk. | | | reaction network | | | |
|---|---|---|---|---|---|---|---|---|---|---|---|---|---|---|
| | | $M$ | $\Delta$ | $c$ | $M$ | $\Delta$ | $c$ | $M$ | $\Delta$ | $c$ | $N$ | $M$ | $\Delta$ | $c$ |
| human | 2310 | 6785 | 0.118 | 9 | 4791 | 0.123 | 0 | 10,798 | 0.220 | 12 | 2303 | 609,377 | 0.236 | 3 |
| human (BiGG) | 1507 | 6527 | 0.197 | 7 | 5003 | 0.142 | 6 | 9601 | 0.245 | 14 | 3311 | 1,307,061 | 0.405 | 2 |
| S. cerevisiae (KEGG) | 1577 | 4426 | 0.132 | 2 | 3135 | 0.139 | 0 | 7155 | 0.203 | 14 | 1449 | 229,483 | 0.235 | 5 |
| S. cerevisiae (BiGG) | 641 | 2693 | 0.160 | 10 | 2093 | 0.101 | 0 | 4181 | 0.269 | 15 | 841 | 153,200 | 0.482 | 5 |
| C. elegans | 1587 | 4250 | 0.144 | 1 | 2991 | 0.122 | 0 | 6899 | 0.218 | 5 | 1378 | 206,869 | 0.203 | 6 |
| M. musculus | 2255 | 6578 | 0.141 | 8 | 4635 | 0.123 | 0 | 10,477 | 0.219 | 14 | 2214 | 562,374 | 0.231 | 7 |
| R. norvegicus | 2098 | 5922 | 0.140 | 1 | 4157 | 0.123 | 0 | 9454 | 0.207 | 9 | 1956 | 459,000 | 0.213 | 7 |
| D. melanogaster | 1897 | 5273 | 0.141 | 2 | 3721 | 0.123 | 0 | 8482 | 0.220 | 10 | 1781 | 361,124 | 0.220 | 5 |
| M. genitalium | 473 | 1030 | 0.184 | 3 | 702 | 0.171 | 0 | 1694 | 0.303 | 4 | 328 | 7104 | 0.381 | 0 |
| E. coli | 1864 | 5262 | 0.127 | 3 | 3639 | 0.021 | 0 | 8406 | 0.197 | 6 | 1800 | 330,247 | 0.263 | 5 |

## Why reaction networks are denser than other representations

To sketch an explanation of this phenomenon for the case of substance and reaction networks (the explanations for the other two types are similar) consider a bipartite network with two types of vertices representing reactions or substances, with edges linking substances to reactions in which they participate. In this network, the degrees of reactions are very narrowly distributed (between two and eight in the human network from KEGG), whereas the degrees of substances are large (between one and 836 in the human KEGG-network). Now, consider a vertex $i$ and assume that the network is sparse and locally tree-like (no cycles within two steps from $i$). The expected degree of a vertex in the projected network (the substance network if the vertex is a substance, or reaction network if the vertex is a reaction) is

$$k = \sum_{k'} k' p_{k'} \sum_{K} p_K K(K-1) = \langle k' \rangle (\mu_2(K) - \langle K \rangle) \tag{1}$$

where $k'$ is the degree (in the bipartite representation) of the kind of vertices projected to, and $K$ is the degree of the kind of vertices projected away from. Angular parentheses represent averages and $\mu_2$ symbolizes the second moment. Simply speaking, a vertex of the other type contributes to $k$ by the square of its degree, because the probability of an edge leading to a vertex of degree $K$ is proportional to $K$, and if an edge leads from $i$ to a neighbor of degree $K$, all the $K-1$ other vertices will be attached to $i$ in the projected network. Now, since $K \geq 1$ for both types of vertices, $\mu_2(K)$ is the leading term in Eq. 1. Since the second moment weights high degree vertices (of the opposite type), higher, we understand that the average degree would be much higher for the reaction networks.

# Appendix 2:
# Currency metabolites

## Substance graphs are optimal simple-graph representations of metabolism

Petter Holme and Mikael Huss

In this file, we provide a figure illustrating the change of relative modularity $\Delta$ as vertices are deleted in the algorithm defining currency metabolites. The darker bars indicate the currency metabolites. The figure represents a substance-graph representation of the human network from the KEGG database. After that we list the sets of currency metabolites for all organisms and network types studied. The order (from top to bottom) represents the order in which they were deleted (like moving from left to right in the figure below).

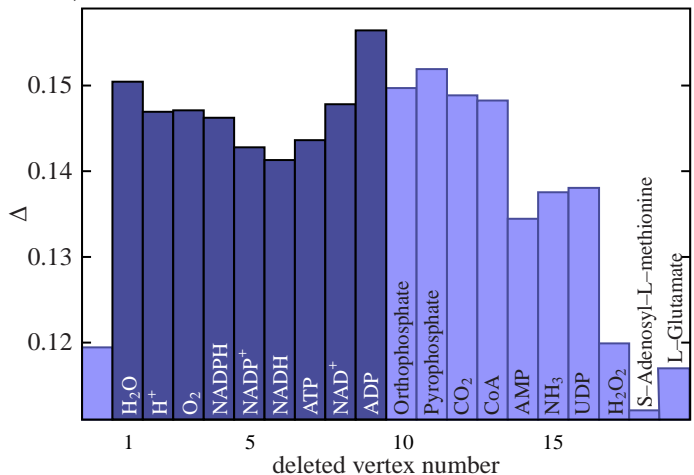

| substr.–prod. | substr.–substr. | substa. | reac. |
|---|---|---|---|
| $H_2O$ | | $H_2O$ | $H_2O$ |
| $H^+$ | | $H^+$ | $H^+$ |
| Oxygen | | Oxygen | ATP |
| NADPH | | $NADP^+$ | |
| $NADP^+$ | | NADPH | |
| ATP | | ATP | |
| NADH | | $NAD^+$ | |
| $NAD^+$ | | NADH | |
| Orthophosphate | | Orthophosphate | |
| | | ADP | |
| | | Pyrophosphate | |
| | | Coenzyme A | |

**Human (KEGG)**

| substr.–prod. | substr.–substr. | substa. | reac. |
|---|---|---|---|
| $H^+$ | $H^+$ | $H^+$ | $H^+$ |
| $H_2O$ | $H_2O$ | $H_2O$ | $H_2O$ |
| Oxygen | ATP | ATP | |
| ATP | Coenzyme A | Oxygen | |
| Orthophosphate | $NADP^+$ | $NADP^+$ | |
| NADPH | Oxygen | NADPH | |
| $NADP^+$ | | Orthophosphate | |
| | | Coenzyme A | |
| | | $NAD^+$ | |
| | | ADP | |
| | | UDP | |
| | | NADH | |
| | | Pyrophosphate | |
| | | $CO_2$ | |

**Human (BiGG)**

| substr.–prod. | substr.–substr. | substa. | reac. |
|---|---|---|---|
| $H_2O$ | | $H_2O$ | $H_2O$ |
| $H^+$ | | $H^+$ | ATP |
| | | NADPH | $H^+$ |
| | | $NADP^+$ | Coenzyme A |
| | | ATP | $CO_2$ |
| | | $NAD^+$ | |
| | | Oxygen | |
| | | NADH | |
| | | Orthophosphate | |
| | | ATP | |
| | | Pyrophosphate | |
| | | Coenzyme A | |
| | | $CO_2$ | |
| | | AMP | |

***S. cerevisiae* KEGG**

| substr.–prod. | substr.–substr. | substa. | reac. |
|---|---|---|---|
| $H^+$ | | $H^+$ | $H^+$ |
| $H_2O$ | | $H_2O$ | $H_2O$ |
| ATP | | ATP | ATP |
| Pyrophosphate | | Pyrophosphate | 2–Oxoglutarate |
| ADP | | ADP | CoA |
| NADPH | | Orthophosphate | |
| Orthophosphate | | $NADP^+$ | |
| $NADP^+$ | | NADPH | |
| AMP | | AMP | |
| $CO_2$ | | $CO_2$ | |
| | | $NAD^+$ | |
| | | NADH | |
| | | CoA | |
| | | Oxygen | |
| | | $NH_4^+$ | |

***S. cerevisiae* BiGG**

| substr.–prod. | substr.–substr. | substa. | reac. |
|---|---|---|---|
| $H_2O$ | | $H_2O$ | $H_2O$ |
| | | $H^+$ | $H^+$ |
| | | Oxygen | ATP |
| | | NADPH | UDP |
| | | $NADP^+$ | $CO_2$ |
| | | | Coenzyme A |

***C. elegans***

| substr.–prod. | substr.–substr. | substa. | reac. |
|---|---|---|---|
| $H_2O$ | | $H_2O$ | $H_2O$ |
| $H^+$ | | $H^+$ | $H^+$ |
| Oxygen | | Oxygen | ATP |
| NADPH | | $NADP^+$ | Coenzyme A |
| $NADP^+$ | | NADPH | Oxygen |
| ATP | | ATP | UDP |
| NADH | | $NAD^+$ | $CO_2$ |
| $NAD^+$ | | NADH | |
| | | Orthophosphate | |
| | | ADP | |
| | | Pyrophosphate | |
| | | Coenzyme A | |
| | | $CO_2$ | |
| | | AMP | |

***M. musculus***

| substr.–prod. | substr.–substr. | substa. | reac. |
|---|---|---|---|
| $H_2O$ | | $H_2O$ | $H_2O$ |
| | | $H^+$ | $H^+$ |
| | | Oxygen | ATP |
| | | $NADP^+$ | Oxygen |
| | | NADPH | Coenzyme A |
| | | $NAD^+$ | UDP |
| | | NADH | $CO_2$ |
| | | ATP | |
| | | Orthophosphate | |

***R. norvegicus***

| substr.–prod. | substr.–substr. | substa. | reac. |
|---|---|---|---|
| $H_2O$ | | $H_2O$ | $H_2O$ |
| $H^+$ | | $H^+$ | $H^+$ |
| | | Oxygen | ATP |
| | | NADPH | Coenzyme A |
| | | $NADP^+$ | $CO_2$ |
| | | $NAD^+$ | |
| | | NADH | |
| | | ATP | |
| | | Orthophosphate | |
| | | ADP | |

**D. melanogaster**

| substr.–prod. | substr.–substr. | substa. | reac. |
|---|---|---|---|
| Pyrophosphate | | ATP | |
| ATP | | $H_2O$ | |
| $H_2O$ | | Pyrophosphate | |
| | | ADP | |

**M. genitalium**

| substr.–prod. | substr.–substr. | substa. | reac. |
|---|---|---|---|
| $H_2O$ | | $H_2O$ | $H_2O$ |
| $H^+$ | | $H^+$ | ATP |
| ATP | | ATP | $H^+$ |
| | | $NAD^+$ | Conezyme A |
| | | NADH | $CO_2$ |
| | | Orthophosphate | |

**E. coli**

# Appendix 3:
## Cluster–Function Maps

# Substance graphs are optimal simple-graph representations of metabolism

## Petter Holme and Mikael Huss

This file contains cluster–function maps for the rest of the organisms (corresponding to Fig. 4 of the paper). All figures are for substance graph representations of the various organisms.

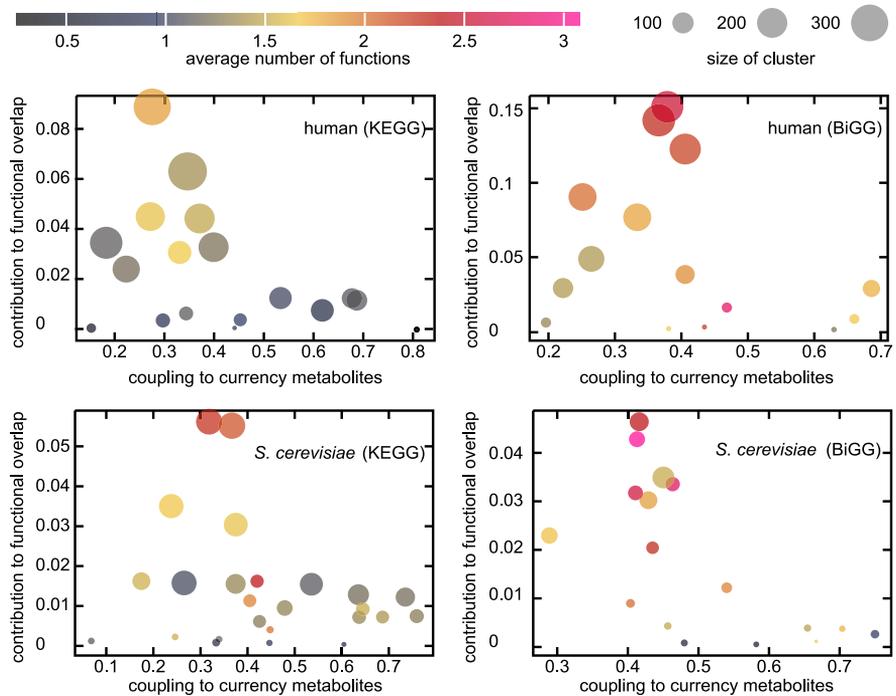

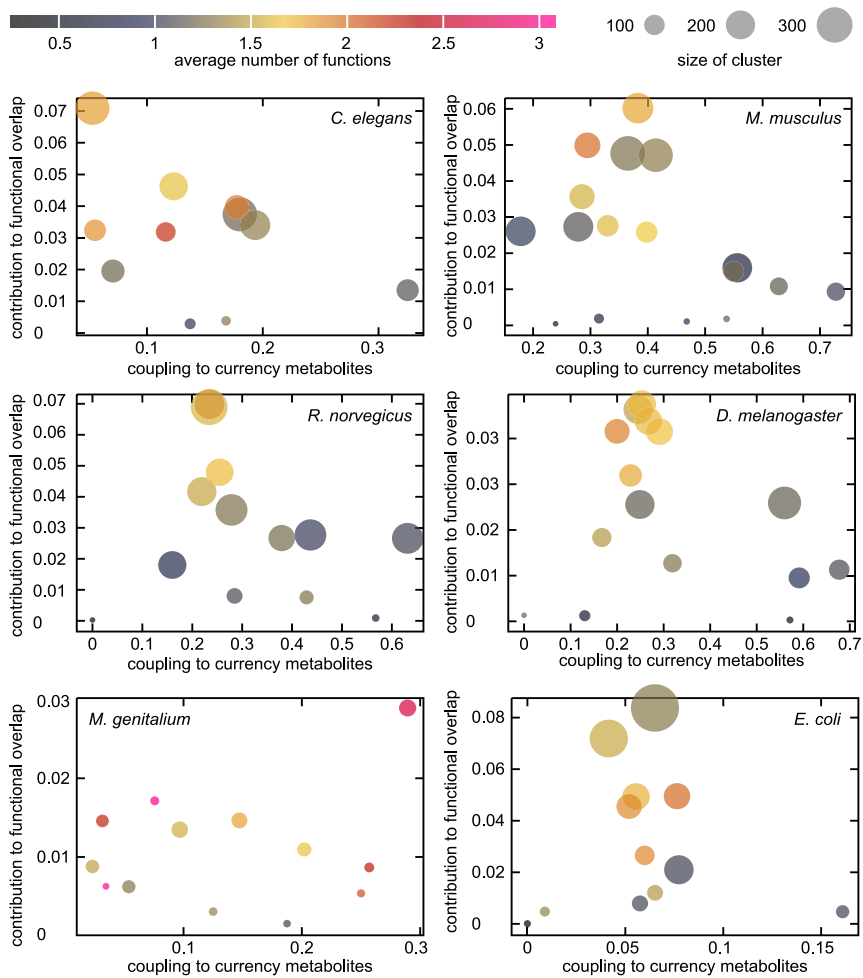

The figure for *M. genitalium* looks a bit different from the picture outlined in the text. Note however that this network does not have any patch with a strong coupling to currency metabolites. It is also an outlier in size, $\Delta$ and other quantities.

# Appendix 1:
# Degree distributions

## Substance graphs are optimal simple-graph representations of metabolism

Petter Holme and Mikael Huss

---

This file contains degree distributions of the studied organisms and the various graph representations. The data is binned logarithmically. Diamonds represent the original degree distribution, including the currency metabolites. Triangles represent the functionally more relevant network, with the currency metabolites deleted.

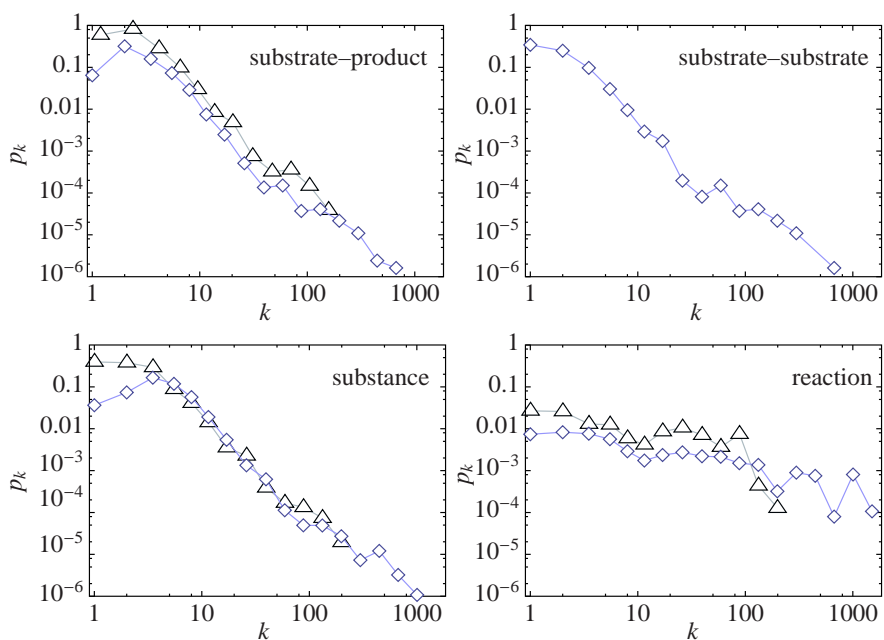

Human (KEGG)

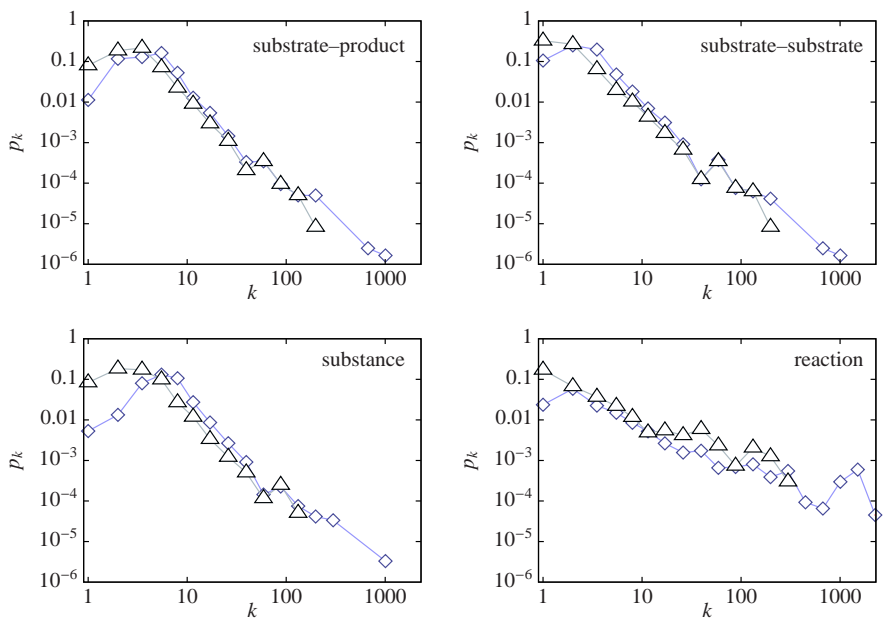

Human (BiGG)

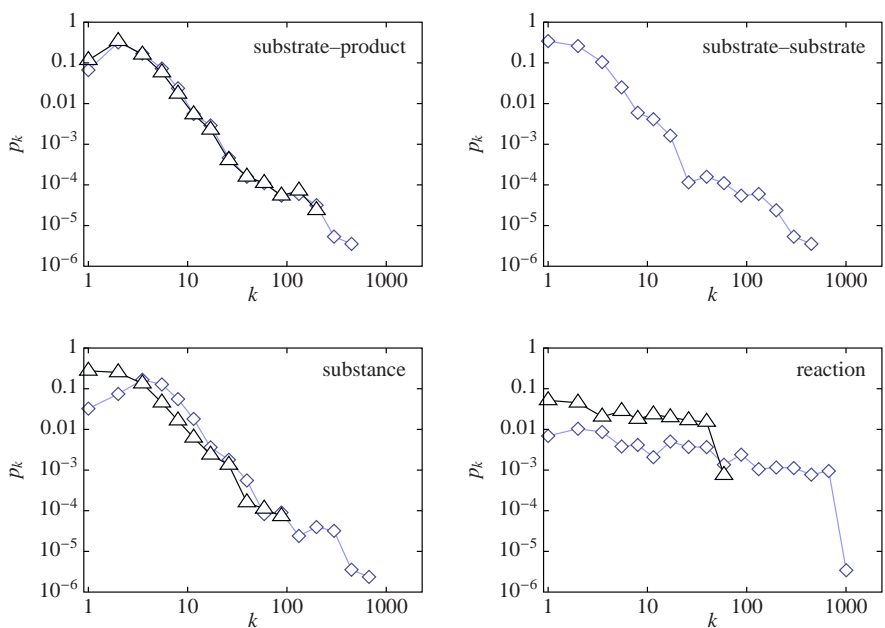

*S. cerevisiae* (BiGG)

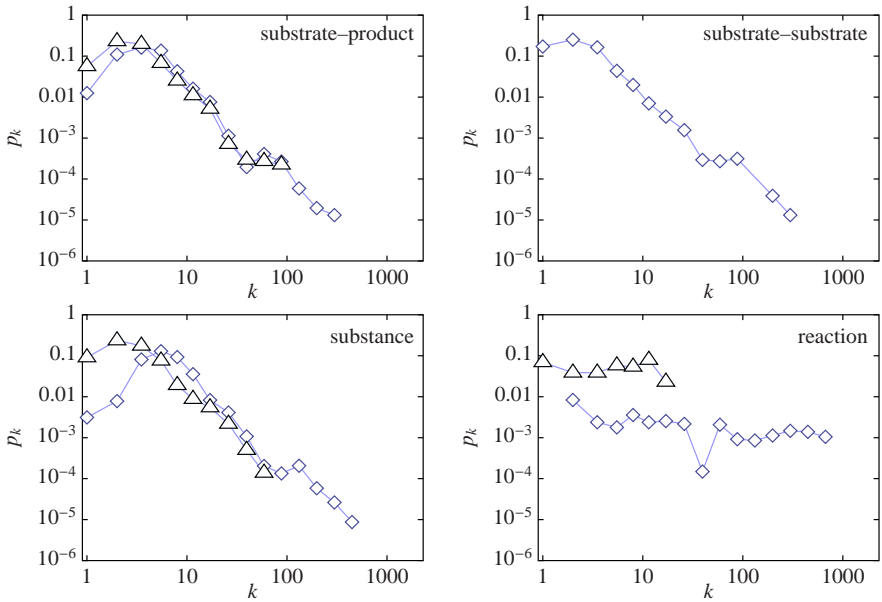

*S. cerevisiae* (BiGG)

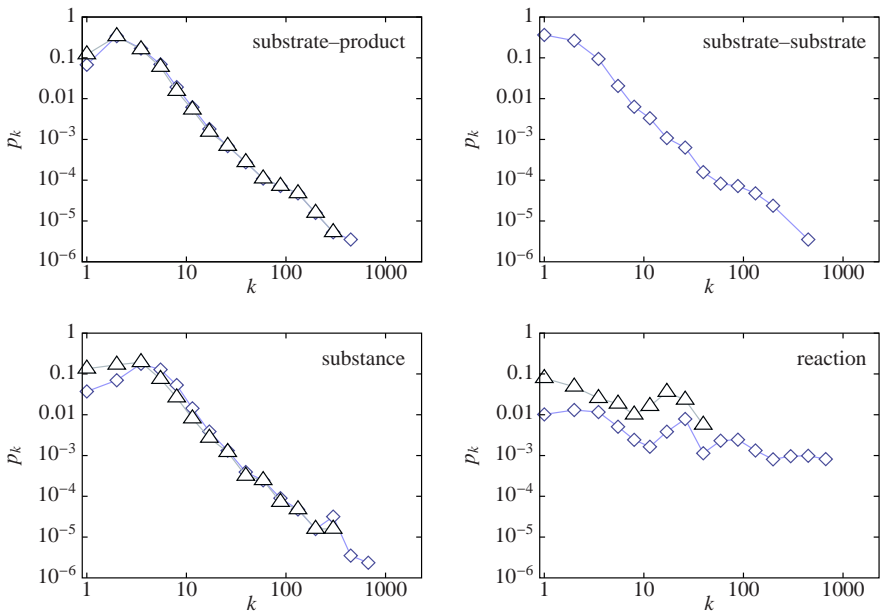

*C. elegans*

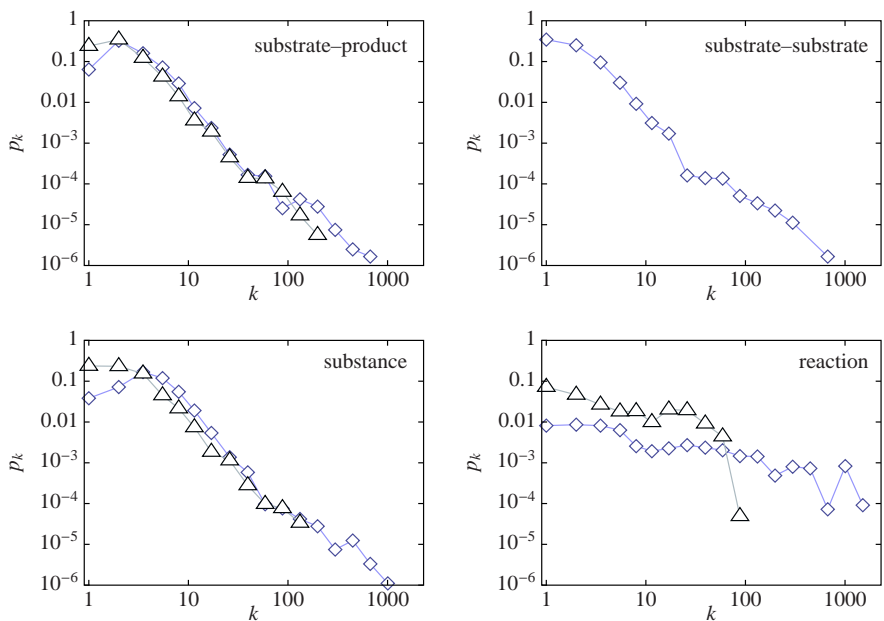

*M. musculus*

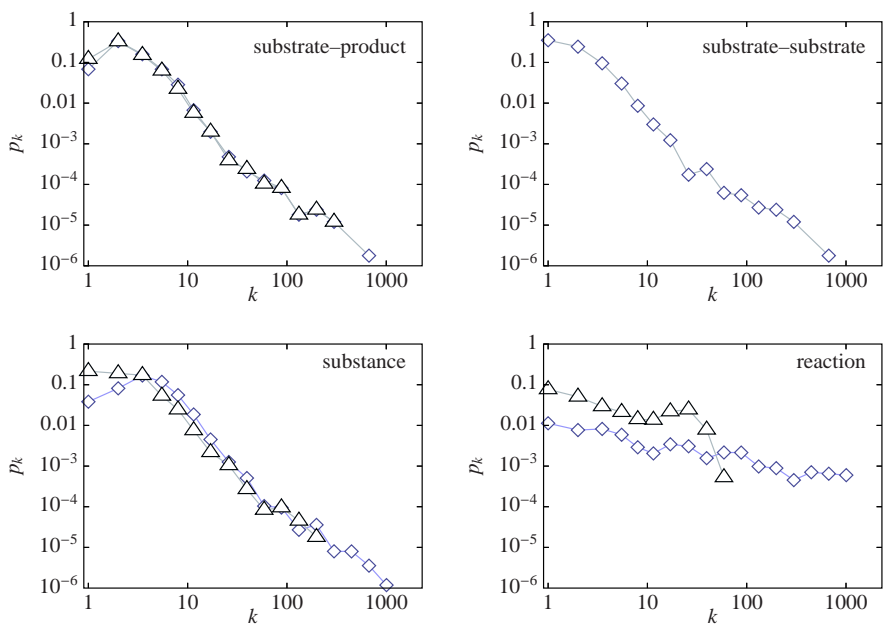

*R. norvegicus*

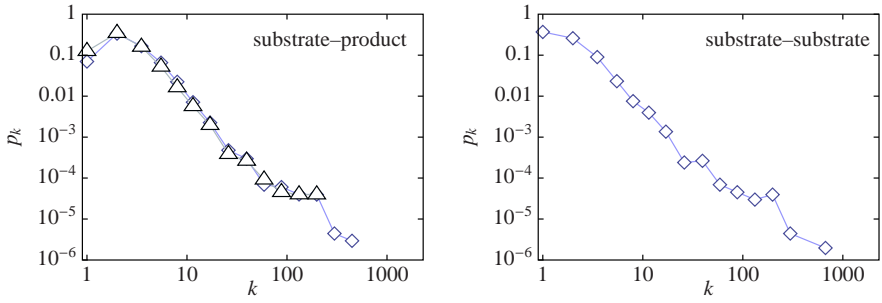

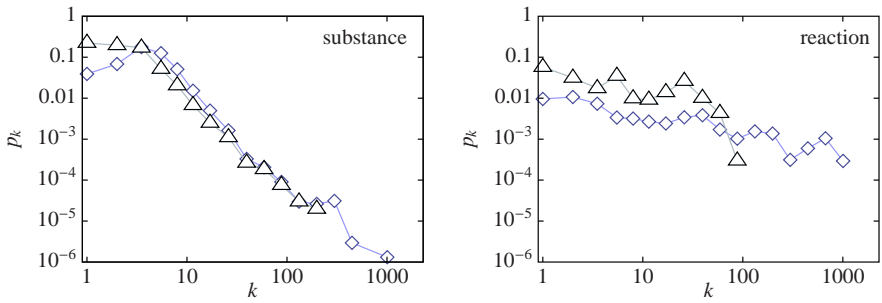

*D. melanogaster*

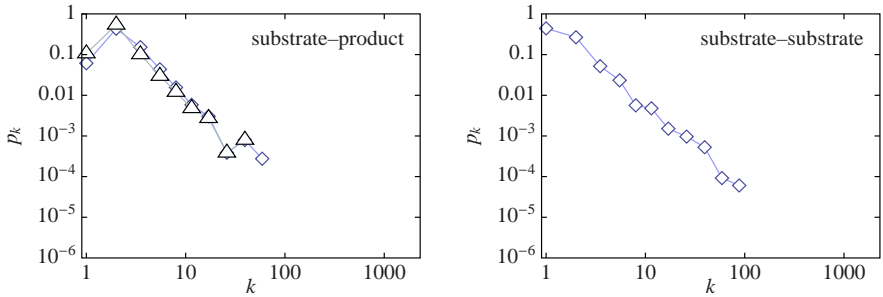

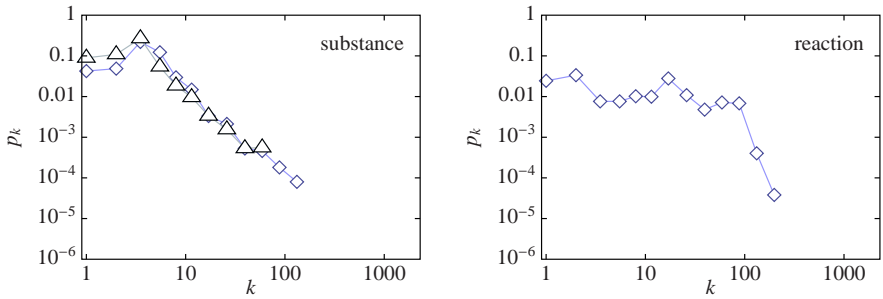

*M. genitalium*